\documentclass[aps,showpacs,preprintnumbers,amsmath,amssymb,superscriptaddress,floatfix,a4paper,nofootinbib,11pt,floatfix,nofootinbib,onecolumn]{revtex4}
\usepackage{graphicx}
\usepackage{bm}
\usepackage{rotating}
\usepackage{array}
\usepackage{xcolor}
\usepackage{amsmath,amssymb}
\usepackage{mathrsfs}
\usepackage{graphicx}
\usepackage{color}
\usepackage{subfigure}
\usepackage{fancyhdr}
\usepackage{multirow}
\usepackage{float}
\usepackage{epsfig}
\usepackage{mathtools}
\usepackage{amsfonts}
\usepackage{bm}
\usepackage{amsmath}
\newcommand{\ba}{\begin{eqnarray}}
\newcommand{\ea}{\end{eqnarray}}
\newtheorem{problem}{Problem}
\begin{document}

\title{\Large \bf Real Classical Geometry with  arbitrary  deficit parameter(s) $\alpha(_{I})$ in Deformed  Jackiw-Teitelboim Gravity}

\author{Davood Momeni} 
\email{davood@squ.edu.om}
\affiliation{Department of Physics, College of Science, Sultan Qaboos University, P.O. Box 36, \\Al-Khodh 123, Muscat, Sultanate of Oman}
%\affiliation{Center for Space Research, North-West University, Mafikeng, South Africa}
%\affiliation{Tomsk State Pedagogical University, TSPU, 634061 Tomsk, Russia}
\begin{abstract}
An interesting deformation of the Jackiw-Teitelboim (JT) gravity has been proposed by Witten by adding a potential term $U(\phi)$ as a self-coupling of the scalar dilaton field. During calculating the path integral over fields, a constraint  comes from integration over $\phi$ as $R(x)+2=2\alpha \delta(\vec{x}-\vec{x}')$. The resulting Euclidean metric suffered from a conical singularity at $\vec{x}=\vec{x}'$. A possible geometry modeled locally in polar coordinates $(r,\varphi)$ by $ds^2=dr^2+r^2d\varphi^2,\varphi \cong \varphi+2\pi-\alpha$. In this letter we showed that there exists another family of "exact" geometries for arbitrary values of the $\alpha$. A pair of exact solutions are found for the case of $\alpha=0$. One represents the static patch of the AdS and the other one is the non static patch of the AdS metric. These solutions were used to construct the Green function for the inhomogeneous model with $\alpha\neq 0$.  We address a type of the phase transition between different patches of the  AdS  in theory because of the discontinuity in the first derivative of the metric at $x=x'$. We extended the study to  the exact space of metrics satisfying the  constraint
$R(x)+2=2\sum_{i=1}^{k}\alpha_i\delta^{(2)}(x-x'_i)$
as a modulo diffeomorphisms for an arbitrary set of the deficit parameters $(\alpha_1,\alpha_2,..,\alpha_k)$. The space is the moduli space of Riemann surfaces of genus $g$ with   $k$ conical singularities located at $x'_k$ denoted by $\mathcal{M}_{g,k}$.
\end{abstract}

%\pacs{Valid PACS appear here}

\maketitle
\date{\today}
%%%%%%%%%%%%%%%%%%%%%%
\section{introduction}
One of the most exciting branches of the current research is finding duality between an exactly solvable lower dimensional quantum system and certain types of the bulk theories with gravity. This duality was inspired by AdS/CFT where at the specific regimes of the coupling, the certain types of the string theories (with gravity) are mapped to strongly coupled large $N$ CFT theories on the flat boundary of the AdS spacetime \cite{Maldacena:1997re}. Depending on the gravitational sector in the bulk, the resultant boundary quantum theory could be different from the CFT one , as well as the bulk could be a non blackhole background. Some basic deformations of AdS/CFT are dualities between asymptotically AdS spaces and QFTs with UV fixed points. For example pure AdS is dual to zero temperature CFTs , blackholes on AdS are duals to finite temperature CFTs while the AdS soliton is dual to QFT with mass gap. The idea of making exact duality between realistic condensed matter systems (exactly solvable) and blackhole physics work initiated with the simple Kitaev model \cite{Kitaev2015} and later its dual bulk theory investigated widely in  \cite{Polchinski:2016xgd,Maldacena:2016hyu}. For certain types of exact solvable Kitaev models, the holographic bulk investigated via gauge/gravity duality. In the Ref.\cite{Saad:2019lba}, the authors  demonstrated the duality between  random Hermitian matrix theory (RMT) and the corresponding  $2d$ bulk theory with  gravity. Gravity in $2d$ is very special and been well studied in past in several works by many authors , see for example \cite{Bagrets:2016cdf}-\cite{Iliesiu:2019xuh} . A reason for such wide study was  firstly its ultraviolet (UV) divergence free form along its simplicity  and secondly its  viable physical interpretations in the context of the string theory. Gravity in $2d$ can be understood as a dilaton gravity as well as a natural reduction of the standard classical general relativity (GR) action from a higher $D$ gravities to $D=2$, see e.g.\cite{Grumiller:2002nm}. Furthermore, Euclidean fully integrable forms of theory in $2d $  were studied in \cite{Bergamin:2004pn}.\par 
JT gravity  is a family of scalar field theories where the scalar field(dilaton) $\phi$ coupled to gravity in two dimensions as a minimal theory of gravity in $2d$ \cite{Jackiw:1984je,Teitelboim:1983ux}. it is possible to remove  UV divergences in two dimensional bulk theory for gravity, as a result it can be considered as a potentially "toy" model for qunatum gravity. A rigorous derivation of the pure JT gravity action can be softly done using a {\it tricky} conformal transformations in $D$ dimensions and then taking the limit of $D\to 2$ \cite{Mann:1992ar}.
\par 
The Euclidean action of the pure JT gravity with a negative cosmological constant  is represented  by 
\begin{eqnarray}\label{jt}
S=-\frac{1}{2}\int_{\Omega} d^2x\sqrt{g}(\phi R+2\phi) + \partial S_{bdy}
\end{eqnarray}
where $\partial S_{bdy}$ stands for any  boundary topological term  (or the usual GHY boundary term) , e.g. Euler characteristic, $R$ is the Ricci scalar curvature of the Euclidean metric tensor $g_{\mu\nu},\mu,\nu=0,1$. By $\Omega$ we mean a region of the spacetime with(without) boundary. It has been demonstrated that the JT gravity action given in (\ref{jt}) enjoys several interesting features from symmetry up to solvability of the equations of the motion (EoMs), see  for example the works done by Refs.\cite{Almheiri:2014cka,Maldacena:2016upp,Engelsoy:2016xyb,Harlow:2018tqv}. If one adds nontrivial couplings between the dilaton and the Abelian $1$-form, the model still shows several physical properties as an exact solvable model
 \cite{Lala:2019inz}. In Ref. \cite{Mertens:2019tcm} the authors studied defects in the JT gravity holographically by  studying  the deformation of the Schwarzian theory as the dual qunatum boundary action. 
 \par
 Very recently a deformation of the pure JT gravity proposed by  
  Witten \cite{Witten:2020wvy}. The model still has RMT dual as a boundary gauge theory in a similar manner as the original JT gravity  \cite{Witten:2020ert}(see \cite{Maxfield:2020ale} also for RMT dual and the relation between critical $3d$ gravity and JT) . Witten investigated a simple deformation of JT gravity by adding a potential term $U(\phi)$ as a self-coupling of the scalar dilaton field. The model reduced to pure JT and furthermore remarkable observation was that the density of energy levels is different from the pure JT. The action for deformed JT (dJT) as proposed 
in Ref.\cite{Witten:2020ert}  takes the form 
\begin{eqnarray}
S=-\frac{1}{2}\int_{\Omega} d^2x\sqrt{g}\Big(\phi R + U(\phi)\Big)\,.\label{ddJT}
\end{eqnarray}
In this work we will study the above model 
as proposed in Ref. \cite{Witten:2020ert} for  $U(\phi)=2\phi + W(\phi)$ , where $W(\phi)$ is 
\begin{eqnarray}
&&W(\phi)=2\sum_{i=1}^{r}\epsilon_{i}\exp\{-\alpha_i \phi\},\ \ \pi<\alpha<2\pi\label{potential}
\end{eqnarray}
To write the above potential function, we assume that the potential function $U(\phi)\sim 2\phi, \ \ \phi\to +\infty$. It is required to get JT theory in the asymptotic limit $\phi\to \infty $. One possible constraint to satisfy the above requirement is to restrict the potential to be as follows,
\begin{eqnarray}\label{delta}
&&\lim_{\phi\to +\infty}|\phi^{1+\delta}(U(\phi)-2\phi)|<1,\ \ \delta \geq 1.
\end{eqnarray}
One possible simple form for such potential function is the one written in expression (\ref{potential}). 
%%%%%%%%%%%%%%%%%%%%%%%%%
A remarkable program for the qunatization of the  JT gravity like theories and viable higher order corrections to it widely studied in the 
in Refs. \cite{Nojiri:2000ja}-\cite{Nojiri:2020tph}.
%%%%%%%%%%%%%%%%%%%%
Some new  exact solutions for dJT studied recently in \cite{Momeni:2020zkx} in favor of the Maldacena's duality conjecture and boundary  Schwarzian theories. In Ref. \cite{Momeni:2020zkx} we showed that how pure AdS   seed metric for pure JT gravity will be deformed in the dJT. In this work we continued our study about dJT. 
The problem we want to address here  is how this perturbative potential  (\ref{potential}) will deform the pure JT gravity bulk geometry. The problem statement will be clearly present 
in Sec.\ref{JT}. It will be formulated on a conformal gauge for Euclidean metric as a nonlinear PDE. 
\par 
The structure of the paper is as follows. In Sec.\ref{JT} we  formulate the problem of the deformed singular metrics with a single deficit parameter $\alpha$.  In Sec.\ref{sec3}, we formulate weak problem via integral balance  technique . In addition, we construct Green function as a solution for the nonlinear PDE formulated in Sec.\ref{dJT}. 
Moreover, in Sec.\ref{nonmini}, we  introduce the phase transition from AdS  to AdS  in dJT. In Sec VI blackhole solutions studied with more than one deficit parameters. In Sec VII a brief discussion given about time dependent geometries with singularity.  We finally conclude our results in the last section.
%%%%%%%%%%%%%%%%%%%%%%
\section{Problem statement }\label{JT}
The Euclidean action for dJT theory eq. (\ref{ddJT}) with potential function (\ref{potential}) for  $\epsilon_{r}=0$ coincides to the pure JT gravity. The aim is to compute the Euclidean path integral (EPI) for first order $\mathcal{O}(\epsilon)$ perturbatively . Following the assumptions made by Witten, it is possible to take the bulk action of order $\epsilon$ for a typical exponential dilatonic potential given by $U(\phi)=2\epsilon e^{-\alpha\phi }$ in the following form
\begin{eqnarray}&&
I=I_{JT}-\epsilon \int \sqrt{g}d^2x e^{-\alpha \phi}\end{eqnarray}
There are higher order terms of $\mathcal{O}(\epsilon^2),n\geq 2$ will be considered if one is interested to see the effects of higher orders. For simplicity we kept the term of order one. The EPI is explicitly written in a perturbative form as the following,
\begin{eqnarray}
&&EPI=\int D\phi Dg \exp\{-I_{JT}\}+\epsilon \int D\phi Dg \exp\{-I_{JT}\}\int d^2 x_1\sqrt{g(x_1)}e^{-\alpha\phi }+\mathcal{O}(\epsilon^2)
\end{eqnarray}
If we only consider the pure JT gravity, the EPI reduces to the partition function of the JT gravity. The first order correction needs an evaluation of an integral in the following form (after a normalization to the volume)
\begin{eqnarray}
&&\int D\phi Dg \exp\{\frac{1}{2}\int d^2x_2\sqrt{g(x_2)}(R+2)\phi(x_2) \}\int d^2 x_1\sqrt{g(x_1)}e^{-\alpha\phi(x_1) }
\end{eqnarray}
The trick to calculate the following integral is to write $\phi(x_2)=\int d^2x_1 \sqrt{g(x_1)} \phi(x_1)\delta(x_1-x_2)$ where the Dirac delta function is defined as
\begin{eqnarray}
&&\int \sqrt{g(x_1)}d^2x_1 \delta(x_1-x_2)=1\label{delta}
\end{eqnarray}
using this field representation, we have
\begin{eqnarray}
\int d^2x_1 \sqrt{g(x_1)} \int D\phi Dg \exp\{\frac{1}{2}\int d^2x_2\sqrt{g(x_2)} \phi(x_2)\Big(R(g(x_2))+2-2\alpha\delta(x_2-x_1)\Big)\}
\end{eqnarray}
We interchange the integration orders, firstly by taking the integral over field $\phi$ ,
\begin{eqnarray}
&&\int Dg \delta(\frac{1}{2}(R(g(x_2))+2-2\alpha\delta(x_2-x_1)))
\end{eqnarray}
The above delta integral can be reduced to a simpler form via the following functional delta function formula,
\begin{eqnarray}
&&\int Dg \delta(f(g))=\int Dg \sum_{i=1}\frac{\delta f(g)}{\delta g}|_{g=g_i}\delta(g-g_i)
\end{eqnarray}
the functional in our case is $f(g)=\frac{1}{2}((R(g(x_2))+2-2\alpha\delta(x_2-x_1))$ and the roots $g_i$ lies on the hypersurface given by $R(g_i(x_2))+2-2\alpha\delta(x_2-x_1)$ in the functional space. Furthermore $\frac{\delta f(g)}{\delta g}|_{g=g_i}=\frac{\delta R}{\delta g}|_{g=g_i}=R^{\mu\nu}|_{g=g_i}\delta g_{\mu\nu}$. Using the above consideration one can complete the partition function steps adequately. The constraint appeared here defines a specific geometry for the metric $g_{\mu\nu}(x_2)$. If one can solve the following partial differential equation (PDE) for $\alpha\neq 0$, we can find the specific geometry for the hypersurface. This is the main task of our paper and we formulate the problem as the following:
\begin{problem}
Find all exact two dimensional  geometries satisfying the constraint equation:\begin{eqnarray}&&R(x)+2=2\alpha \delta (x-x'),\ \  \alpha\neq 0
\label{problem}
\end{eqnarray}
The resulting geometries suffer from a conical singularity at $x=x'$. 
\end{problem}
In the Witten's paper, it has been claimed that the geometry "can be modeled locally" by a conical flat geometry 
\begin{eqnarray}&& ds^2=dr^2+r^2d\varphi^2,\varphi \cong \varphi+2\pi-\alpha
\end{eqnarray}
and later it has been claimed that "there is no real classical geometry " for $\alpha> 2\pi$. In this paper we show that there is classical geometry for any arbitrary value of the $\alpha$. The problem of finding a solution to this problem reduces to constructing proper Green function for a non linear operator. Before we solve the above problem and find "exact" non trivial geometry for it, in the next section we will give a physics to the deficit parameter $\alpha$ via the method of integral balance.
%%%%%%%%%%%%%%%%%%
\section{The integral balance method and the meaning of $\alpha$}\label{sec3}
The method of integral balance is trying to formulate the weak problem for a given general PDE \cite{Yehuda Pinchover}. Indeed, we know that there is a naive connection between an integral balance and the associated differential operator equation. It is worth mentioning here that the method of integral balance is more fundamental and can only be inverted into a PDE form, when the field functions are sufficiently smooth. Because we need the explicit form of the PDE for our problem formulated in the previous section, i.e,  (\ref{problem}) , we adopt an Euclidean two dimensional geometry in the following null coordinates
\begin{eqnarray}
ds^2=e^{\psi(u,v)}dudv,\ \ u=z+t,v=z-t
\end{eqnarray}
here $t$ is the Euclidean time.
An explicit expression for the Ricci scalar is $R=-4e^{-\psi}\partial_u\partial_v\psi$. Because we have to satisfy the normalization condition given in eq. (\ref{delta}), we multiply both sides of the equation (\ref{problem}) by factor $\sqrt{g}=\frac{1}{2}e^{\psi(u,v)}$, by dropping the factor $2$, we obtain the following nonlinear PDE,
\begin{eqnarray}
&& \underbrace{\frac{e^{\psi}}{2}}_{\sqrt{g}}
-\partial_u\partial_v\psi=\alpha \underbrace{\frac{e^{\psi}}{2}}_{\sqrt{g}} \delta(u-u')\delta(v-v')\label{PDE1}
\end{eqnarray}
Now the Problem simplifies to find a non trivial solution for the non linear inhomogeneous PDE eq. (\ref{PDE1}). Actually the solution is nothing but  the Green equation for any "arbitrary value of the deficit parameter $\alpha$. We use the term of Green equation and consequently the non trivial solution for the above PDE is basically a Green function defined in the following form
\begin{eqnarray}
&& \hat{O} G(x|x')= \frac{\alpha}{2}\exp\{G(x|x')\} \delta(u-u')\delta(v-v')\label{green1}
\end{eqnarray}
where the non linear differential operator $\hat{O}$ is defined as
\begin{eqnarray}
&&\hat{O}[..]\equiv \frac{e^{[..]}}{2}-\partial_u\partial_v[..]\label{O}
\end{eqnarray} 
and $\psi\equiv G(x|x'), \ \ x\equiv (u,v)$. Although the operator is not Hermitian (we will check it later) or linear but we are very lucky to have at least two exact solutions for it. One exact solution for the operator found in Ref.\cite{Momeni:2020zkx}. That solution corresponds to the pure AdS metric written in the null coordinates. The other solution as we will show in the next section will represent another  AdS solution but in the non static patch (cosmological patch). Both solutions are exact solutions and will be used effectively to find Green function in the next section. Solving the above PDE is our plan in the next section.\par
Let us see whether the operator $\hat{O}$ is self adjoint on the domain $\Omega_2 $ as compact version of the real domain $\Omega_1$.
We know that the usual Euclidean AdS (half plane) coordinates (hyperbolic geometry) are living in the following domain
 \begin{eqnarray}
 &&\Omega_1=\{z,t\in (0,\infty)\times (-\infty,+\infty)\}
 \end{eqnarray}
 the above non compact domain will be mapped to another non compact null domain,
 \begin{eqnarray}
 &&\Omega_2=\{u,v\in(u_{-},u_{+})\times(v_{-},v_{+})\}
 \end{eqnarray}
 basically $|u_{\pm}|,|v_{\pm}|\to \infty $ but we kept it as some types of the "conformal boundaries" for the new mapping domain which actually live in very far away region. We know that the operator is self-adjoint operator (or equivalently, a Hermitian operator) if and only if  for a pair of the functions, it satisfies the following integral ,
 \begin{eqnarray}
 &&\int_{\Omega} \sqrt{g}dudv \Phi_1^{*}(u,v)\hat{O}\Phi_2(u,v)=
 \int_{\Omega} \sqrt{g}dudv \Big(\hat{O}\Phi_1(u,v)\Big)^{*}\Phi_2(u,v)\label{selfadjoint}
 \end{eqnarray}
We simply use the usual definition of the self-adjoint operator on a Hilbert space in the finite-dimensional space, the only difference backs to the integral measure, instead of the flat space we use the covariant volume element. It is illustrative to mention here that the partial derivative term in the (\ref{O}), looks just like a kinetic energy operator, consequently is a Hermitian term , while  the first exponential part doesn't satisfy the (\ref{selfadjoint}) conditions. As a result the operator is non Hermitian (not self-adjoint). But at the end of day, still the operator enjoys some linearity at the level of solutions. That means for a pair of exact solution for the operator, i.e, the kernel pair functions $\psi_{1,2}(u,v)$,
\begin{eqnarray}
&&\hat{O}\psi_{1,2}(u,v)=0\label{kernel}
\end{eqnarray}
one can show that the solutions remain linear independent even when the operator is nonlinear by itself. In the Language of the PDEs, the partial Wronskian  of the functions is non zero. The partial Wronskian for solutions, $W_u(\psi_1,\psi_2)$ is  
\begin{eqnarray}
&&W_u(\psi_1,\psi_2)=\psi_1\partial_u\psi_2-\psi_2\partial_u\psi_1
\end{eqnarray}
We will prove it later in next section (\ref{dJT}).
 In this section we only focus on studying a special discrete regime of the above PDE, i.e, the integral balance technique.  We integrate the PDE (\ref{PDE1}) from the lower to the upper boundaries, we obtain,
\begin{eqnarray}
&&\underbrace{\frac{1}{2}\int_{v_{-}}^{v_{+}}dv \int_{u_{-}}^{u_{+}}e^{\psi}du}_\text{Entirely space volume} -\int_{v_{-}}^{v_{+}}dv \int_{u_{-}}^{u_{+}}du\partial_u\partial_v\psi=\alpha \underbrace{\int_{v_{-}}^{v_{+}}dv \int_{u_{-}}^{u_{+}} \frac{1 }{2} \delta(u-u')\delta(v-v')e^{\psi}du}_{1}
\label{balance1}
\end{eqnarray}
We assume that $u_{-}<u'<u_{+},v_{-}<v'<v_{+}$, although the $\leq $ also satisfies our arguments, but it is too risky to work with integral of the distribution functions over domains where the initial or final points are singularities of the integrand. As we understood $\psi$ is just the Green function, and it always remains continuous as well as symmetric respecting the exchange of $x\to x'$ and with a discontinuity at the point $x=x'$. Using the mean value theorem, the first term of the relation (\ref{balance1}) is the total volume of the spacetime manifold, i.e, $V_{tot}$, the right hand side is just simply $\alpha$ (remembering normalization condition (\ref{delta})). The second term can be integrated part by part carefully and finally we obtain
\begin{eqnarray}
&&\alpha=V_{tot}-\psi(u_{+},v_{+})+\psi(u_{+},v_{-})+\psi(u_{-},v_{+})-\psi(u_{-},v_{-})\label{PDE3}
\end{eqnarray}
The relation (\ref{PDE3}) is considered as weak problem version of the original (\ref{problem}). In particular it is suitable to integrate numerically and find the metric profile function $\psi$. The physical meaning of the $\alpha$ is encoded in (\ref{PDE3}): the deficit parameter $\alpha$ is the particular volume of the spacetime by excluding the upper $\psi(u_{+},v_{+})$ and lower $\psi(u_{-},v_{-})$ values of the metric function and including the corner sides. We now demonstrate the construction of a weak solution $\psi(u,v) $ that is a continuously differentiable function over the whole plane. The only exception is for discontinuities along a curve $u=\gamma(v)$. Because the solution is smooth on both sides of $\gamma$ , it is easy to show that it satisfies the PDE (\ref{PDE1}). We write the weak solution in the following form
\begin{eqnarray}
&&\psi(u_{-},u_{+}|v)=\int_{u_{-}}^{u_{+}} du \psi(\zeta,v)d\zeta\end{eqnarray}By plugging it to the (\ref{PDE1}) and integration we obtain
\begin{eqnarray}
&&\frac{1}{2}\exp\{\int_{u_{-}}^{u_{+}} du \psi(\zeta,v)d\zeta\}-\partial_v(\psi(u_{+},v)-\psi(u_{-},v))=\alpha \label{weak1}
\end{eqnarray}
We need to compute $\gamma$. For this purpose we write the weak formulation (\ref{weak1}) in the form
\begin{eqnarray}
&&\frac{1}{2}\exp\{\int_{u_{-}}^{\gamma(v)} \psi(\zeta,v)d\zeta+\int_{\gamma(v)}^{u_{+}} \psi(\zeta,v)d\zeta\}-\partial_v(\psi(u_{+},v)-\psi(u_{-},v))=\alpha \label{weak}
\end{eqnarray}
Differentiating the above integrals with respect to $\gamma$ and using the PDE itself and after performing the integration we can show that the curve $\gamma$ propagates  at uniform speed. The speed is the given by
\begin{eqnarray}
&&\frac{dv}{d\gamma}=\frac{\partial_v\psi(u_{+},v)-\partial_v\psi(u_{-},v)}{ \partial_v \psi(u_{-},u_{+}|v)}
\end{eqnarray}
the average of the propagation speeds on the left and right ends. After this interpretation, we go back to solving the PDE (\ref{PDE1}) and finding the metric functions which are associated with it.

%%%%%%%%%%%%%%%%%%%%
\section{ Constructing Green function}\label{dJT}
The metric function $\psi(x,x')$ defines Green functions for the non linear differential operator $\mathcal{O}$. Explicitly we observe that there are a pair of exact solutions for the homogeneous case, $\mathcal{O}\psi_{1,2}=0$ given as the following
\begin{eqnarray}
\label{trial}&&\psi_1(x)=2\log (2|v+u|^{-1}),\ \ \psi_2(x)=2\log (2|v-u|^{-1}).
\end{eqnarray}
These solutions are found using a direct ansatz for the solutions  $\psi\sim|u\pm v|^n$. After substituting into the field equation (\ref{kernel}), we find $n=-2$ for both cases $\psi_{1,2}$. The proportionality constant can be fitted after a more investigation of the homogeneous PDE. In the above solutions (whose are not the unique solutions because of the non linearity), the first $\psi_1$ is suitable for AdS boundary regimes $z\to 0$ while the second one works for non static cosmological AdS spacetime and suitable for regions with $z\to \infty$ (spatial boundary regions). Using the above pair of the "exact" modes we can show that the following symmetric preposition is a suggestion for Green function,
%how to typeset multi-part definitions in LaTeX.
\begin{eqnarray}
\label{green33}
G(x|x') =\left\{	\begin{array}{ll}	C \psi_1(u,v)\psi_2(u',v')  & \mbox{if } u < u', v<v' \\			C \psi_1(u',v')\psi_2(u,v) & \mbox{if } u > u', v>v'	
\end{array}\right.
\end{eqnarray}
We kept the expression for Green function in terms of any pair of exact modes, including the hypothetical solutions we presented  in expressions (\ref{trial}). To fix the parameter $C$, one needs to integrate both sides of the inhomogeneous PDE (\ref{green1}). in a domain close to the singularity point, $x=x'$, we assume that the Green function remains continuous on the singularity, it provides the following limiting integral constraint via the mean value theorem in calculus,
\begin{eqnarray}&&
\lim _{\epsilon_a\to 0}\int_{v'-\epsilon_2}^{v'+\epsilon_2}dv\int _{u'-\epsilon_1}^{u'+\epsilon_1} du \sqrt{g}=0.
\end{eqnarray}
The above integral denotes shrinking of the volume enclosed by the singularity point $x=x'$.
The other terms in the integration process can be calculated easily, after a simple usage of the Leibniz formula we obtain,
\begin{eqnarray}\label{green22}
&&\psi_1(u',v')\Big(\psi_2(u'+\epsilon_1,v'+\epsilon_2) -\psi_2(u'-\epsilon_1,v'+\epsilon_2)\Big)\\&&\nonumber-\psi_2(u',v')\Big(\psi_1(u'+\epsilon_1,v'-\epsilon_2)-\psi_1(u'-\epsilon_1,v'-\epsilon_2)\Big)=-C^{-1}
\end{eqnarray}
We apply the Taylor series by assuming that $|\epsilon_a|\ll |x'|$, as the following,
\begin{eqnarray}
&&\psi_a(u'\pm \epsilon_1,v'\pm \epsilon_2)=\psi_a(u',v')\pm \epsilon_1\partial_u\psi_a|_{u',v'}\pm \epsilon_2\partial_v\psi_a|_{u',v'}+...
\end{eqnarray}
we simplify the (\ref{green22}) to the following,
\begin{eqnarray}
&&C=-\frac{\alpha'}{2W_u(\psi_1,\psi_2)|_{u',v'}}
\end{eqnarray}
and we urge to redefine the deficit angle $\alpha'\equiv \frac{\alpha}{\epsilon_1}$. By inserting $C$ into the multi-part Green function (\ref{green33}) we finally obtain the metric function associate to the (\ref{problem}),
\begin{eqnarray}
&&\psi(x,x')=-\frac{\alpha'}{2W_u(\psi_1,\psi_2)|_{u',v'}}\psi_1(u_{<},v_{<})\psi_2(u_{>},v_{>})
\end{eqnarray}
here $x_{<},x_{>} $ refers to $x,x'$. A non trivial fully classical metric with deficit parameter can be written in the following explicit form
\begin{eqnarray}
&&ds^2=\exp\{-\frac{\alpha'}{2W_u(\psi_1,\psi_2)|_{u',v'}}\psi_1(u_{<},v_{<})\psi_2(u_{>},v_{>})\}dudv
\end{eqnarray}
Using the pair of the trial solutions (\ref{trial}) one can show that,
\begin{eqnarray}
&&W_u(\psi_1,\psi_2)=\psi_1\exp\{\frac{\psi_2}{2}\}+\psi_2\exp\{\frac{\psi_1}{2}\}
\end{eqnarray}
we end out with a smoothly (but with discontinuity at the singularity point) multi-part metric,%\psi_1(x)=2\log (2|v+u|^{-1}),\ \ \psi_1(x)=2\log (2|v-u|^{-1})%
\begin{eqnarray}
\label{green3}
ds^2 =dudv\left\{	\begin{array}{ll}\exp\{-\Big(\frac{\alpha(v'-u')}{2\epsilon_1\log(\frac{v'-u'}{v'+u'})}\Big)\log (2|v+u|^{-1})\log (2|v'-u'|^{-1})\}	  & \mbox{if } u < u', v<v' \\\exp\{-\Big(\frac{\alpha(v'-u')}{2\epsilon_1\log(\frac{v'-u'}{v'+u'})}\Big)	\log (2|v-u|^{-1})\log (2|v'+u'|^{-1})\}	 & \mbox{if } u > u', v>v'	\end{array}\right.
\end{eqnarray}
The metric is continuous at boundary  $x=x'$
\begin{eqnarray}
&&g^{>}_{\mu\nu}|_{x=x'}=g^{<}_{\mu\nu}|_{x=x'}.
\end{eqnarray}
but there is a discontinuity in the first derivative coming from the Green function ,
\begin{eqnarray}
&&\partial_{u}g^{>}_{\mu\nu}|_{x=x'}\neq \partial_{u}g^{<}_{\mu\nu}|_{x=x'}.
\end{eqnarray}
the last can be interpreted as a discontinuity in the affine connection for the spacetime. The reason is that the non vanishing components for the Christoffel symbol are given either as by $\partial_{u}\psi$ or $\partial_{v}\psi$. Note that the Green functions  have  discontinuity (or jumping) for both  $\partial_{u,v}$. The discontinuity which is located at $\partial_{u}G(x|x')$ is proportional (more precisely equals) the Christoffel symbol $\Gamma^{t}_{\mu\nu}$. Any discontinuity in the $\Gamma$ will be transferred directly to the geodesics of the test particle. Basically we guess the trajectory of a test particle undergoes a critically. 
A more interesting interpretation for the discontinuity will be presented in the nest section where we will address a phase transition between metrics which are representing the trial solutions $\psi_1,\psi_2$.
%%%%%%%%%%%%%%%%%%%%%
\section{More about the metrics associated to $\psi_1,\psi_2$ }\label{nonmini}
The trial solutions given in expressions (\ref{trial}) define two geometries with very interesting features,
\begin{eqnarray}&&
ds_1^2=\frac{4dudv}{(u+v)^2}, ,\\&& ds_2^2=\frac{4dudv}{(v-u)^2}.
\end{eqnarray}
The Ricci scalar is $R=-2$ for both metrics except for asymptotic regions $u\to \pm v$ (conformal boundaries). If one write the metrics in the standard Poincare's coordinates, it represents a half plane metric for $z>0$. We mention here that any other solution except the AdS is a quotient $AdS/\Gamma$, here $\Gamma$ denotes a discrete subgroup of $SL(2,\mathcal{R})$ and as a result it is not unique solution. 
It is instructive to rewrite the above metrics in the usual Poincare coordinates (not the Euclidean) where $u=z+i\tau,v=z-i\tau$,one immediately find that the metrics corresponding to the $\psi_1$ and $\psi_2$ are both represent $AdS_2$,
\begin{eqnarray}
&&ds_1^2=\frac{dz^2+d\tau^2}{z^2},\ \  AdS,\\&& ds_2^2=-\frac{dz^2+d\tau^2}{\tau^2}, \ \  AdS.
\end{eqnarray}
The first can be interpreted as $AdS_2$ in the static path including the AdS boundary , the second one after a signature change represents non  static patch of the AdS (if such patch existed at all). For the first metric, the dual system is supposed to lie on the conformal boundary $z=0$. Technically the conformal boundary locates at infinitely far away region. 
The signature change from first to the second metric shows that the trial solutions belong to different metrics. That implies that, although the PDE for these functions is nonlinear but somehow "non linear" Independence of the solutions still remains valid. Although one can easily show that the AdS metric with wrong signature (signature changed) can be transformed to the AdS metric (both solutions are obtained  as  exact solutions for the homogeneous case with $\alpha=0$),  if we let the metric coordinates $(z,\tau)$  undergo a complex conformal transformations ,
\begin{eqnarray}
&&(z\to i\tau, \tau\to  z) \Longrightarrow (z\to iz, \tau\to\tau)  \\&&
ds_2^2=-\frac{dz^2+d\tau^2}{z^2}
\end{eqnarray}
The new metric is considered as the standard AdS which undergoes a signature change. Note that dJT gravity is diffeomorphism invariance  as well as conformal invariance theory, i.e, any signature change of the metric $g_{\mu\nu}\to -g_{\mu\nu}$ doesn't change the action. 
The reason is that under such transformation, the Ricci scalar remains unchanged because  $R=g_{\mu\nu}R^{\mu\nu}\to R$. But signature change probably makes difference for the dilaton profile. Under such transformation, the dilaton probably will change. 
\par
The signature change from static $AdS$ to the non static patch $AdS $, or equivalently from $ds_1^2\to  ds_2^2$ can be understood by studying the trajectories of a test  particle in the background of both metrics. For this purpose we have to write down the set of the equations of the motion (EoM) for trajectory, for simplicity we just consider a photon path, basically with a suitable parametrization, we can find the trajectories $(\tau(\zeta),z(\zeta))$ by minimization of the following string like actions,
\begin{eqnarray}
&&S_1=\pm \int\frac{d\zeta}{z}\sqrt{\dot{z}^2+\dot\tau^2},\ \ \mbox{for}\ \  ds_1^2,\\&&S_2=\pm \int\frac{d\zeta}{\tau}\sqrt{\dot{z}^2+\dot\tau^2},\ \ \mbox{for}\ \  ds_2^2
\end{eqnarray}
for both action functionals there are a pair of the conserved charges, read as
\begin{eqnarray}
\label{es}
&&E_1=\frac{\dot\tau}{z\sqrt{\dot{z}^2+\dot\tau^2}},\ \ E_2=\frac{\dot z}{\tau \sqrt{\dot{z}^2+\dot\tau^2}}
\end{eqnarray}
Here $E_1$ has the meaning of an energy while $E_2$ defines a conserved transnational momentum along the $z$ coordinate.
one can easily show that the above first integrals are the unique EoMs for the trajectories , the second EoM trivially satisfied. By using the conserved charges (\ref{es}), we can show that the trajectory of the test particle reduces to the existed points on the Hamilton surface (first integral), 
\begin{eqnarray}
&& \tau^2-(\frac{E_1}{E_2})z^2=b
\end{eqnarray}
here $b$ is constant. Depending on the sign of this parameter, we have the following cases,
\begin{itemize}
\item If $b>0,\frac{E_1}{E_2}<0$, then the trajectory is either an elliptic (for $E_1\neq E_2$ ) or circular (for  $E_1= E_2$ ). The trajectory corresponds to non static patch of the AdS. 
\item If $b\geq 0,\frac{E_1}{E_2}>0$, then the trajectory is either a hyperbolic  (for $E_1\neq E_2$ ) or line (for  $E_1= E_2,b=0$ ), its simply the typical trajectory in static patch of the  AdS. 
\item For $b<0$, one can show that the situation remains the same as the above cases if one performs the coordinates transformations as $\tau\to z\sqrt{|\frac{E_1}{E_2}|} $. 
\end{itemize}
The signature change can be understood as a change in the photon trajectory from static region in the AdS to non static (cosmological) patch adequately. If the conserved charges have different signatures , then the trajectory is closed, it corresponds to $AdS$ as a possibility to have minimal surfaces in hyperbolic spaces with negative curvature (that is AdS). For the case when the conserved changes have different signs, the particle trajectory falls down into non static patch (or equivalently into an open dS universe) , there is no minimal surface that corresponds to the non static patch of the AdS. We can understand the transition from AdS to AdS as a phase transition. It corresponds to the Green function we obtained . Indeed as we know, the derivative of the Green function has a jump discontinuity at $x=x'$. This type of discontinuity is interpreted as a discontinuity in the first derivative of the metric or more precisely the Christoffel symbols. Let us explain it in a more concrete way: The Green function as the response function of the inhomogeneous Witten's equation is dual to the two point function of a boundary operator $\hat{O}$. In our case it is related to the probability of measuring a field $\phi(x)$ when the source is at $x'$. A discontinuity in the metric function implies a jumping in the free energy of the system. It is known that the free energy can be written in terms of the thermal Green functions for a bulk/boundary theory. Remembering that the Green functions is the two point correlation function and expectation value of the operators can be expressed in terms of the partition function using the path integrals. We postulate that the following discontinuity in the metric for a non zero deficit parameter $\alpha$ addresses the phase transition from static to cosmological patch of the AdS  in dJT at least in a formal form. For a better understanding of the phase transition one should compare the free energy for bulk theory in dJT for different patches of the AdS. We guess that a more careful calculation will support the argument which we stated here . 
%%%%%%%%%%%%%%%%%%%%%%
\section{Black hole solutions }
The Green function metric derived in the previous section and it's interesting phase transition scenario can be recast into a spherically symmetric Euclidean metric in the Schwarzschild coordinates $x^{\mu}=(t,r)$. As we learned from study of the exact solutions in the null coordinates ,to define the metric for the geometry we need to specify only one gauge function here could be function of  $(r)$ as the following,
\begin{eqnarray}
&&ds^2=A(t)dt^2+\frac{dr^2}{A(t)}\label{metric22}
\end{eqnarray}
Any other representation of the metric with two arbitrary functions, for example $A(r),B(r)$ can be recast to the above case after a suitable reparametrization of the radial(spatial)   coordinate $r$. For the case of time dependent metric it is very difficult to reduce the metric to the null coordinates form or eliminate one of the metrics functions. A reason is that one can't deduce a simple Birkhoff's theorem for JT gravity. It is very hard to prove that the JT gravity posses only   static and asymptotically flat solutions in the absence of any other matter field contents. We don't study the validity of such fundamental theorem in the JT gravity as also we are not sure about such proof in more general cases of the UV free two dimensional theories for quantum gravity.  \par
If we limit our study to the case where metric remains time independent, it is easy to show that for non singular case , i.e, when $\alpha=0$, the classical field equations provides a class of the  exact solutions for the metric found in \cite{Witten:2020ert}. Furthermore for a general class of the deformation potentials $U(\phi)$, the metric function given by 
\begin{eqnarray}
&&A(r)=\int_{r_h}^{r}U(r')dr'
\end{eqnarray}
in the above suggested solution, we supposed that there exists a blackhole solution with a null hyperbolic surface (horizon) which is located at  $r=r_h$ and $A(r_h)=0$. The value of the dilaton field $\phi(r_h)=\phi_h$ kept finite and positive. One can show that the near horizon geometry of the metric is thermal region with  the temperature given by 
\begin{eqnarray}
&&T=\frac{U(\phi_h)}{4\pi}
\end{eqnarray}
If one try to satisfy the JT gravity asymptotically bound on the potentail function $U(\phi)$ defined in expression (\ref{delta}), we can show that the asymptotic form for the metric at the limit $\phi\to \infty$ is 
\begin{eqnarray}
&&A(r)=r^2-b+\mathcal{O}(r^{-\delta })
\end{eqnarray}
Using simple arguments, one can show that the first law of the thermodynamics still holds if one identify the parameter $b$ as a quasi energy $E=\frac{b}{2}$ defined via the Gibbons-Hawking-York surface.
%%%%%%%%%%%%%%%%%%%%%%%%%%%%%%%%%%%%
\par 
Our aim in this section is to find 
 an exact solution for the Witten's equation (\ref{problem}) with a single deficit angle with metric form given by (\ref{metric22}). A remarkable observation is that the metric keeps the Dirac function same as the flat Euclidean metric. That is because the metric is uni-modular  i.e, $\det{g_{\mu\nu}}=1$,the normalization condition for Dirac delta function  
 \begin{eqnarray}
&&\int \sqrt{g(x_1)}d^2x_1 \delta(x_1-x_2)=\int_{M^2} dr dt \delta(r-r')\delta(t-t')=1
\end{eqnarray}
 Using the spherical symmetry of the metric one can 
 write the Dirac delta function only as a $r$ dependent form
 \begin{eqnarray}
 &&\delta(x-x')=\delta(r-r')
 \end{eqnarray}
 where we implied the time Independence of the metric as well as spherical symmetry. Using the metric, 
we can compute the Ricci scalar as $R=-A''$, the equation for the modular space reduces to the following linear ODE:
\begin{eqnarray}\label{ODEA}
&&A''=2(1-\alpha \delta(r-r')),\ \ \alpha\neq 0.
\end{eqnarray}
for single deficit parameter $\alpha$, and a single conical singularity located at $r=r'$.  An exact solution for the metric can be obtained using simple integration techniques,
\begin{eqnarray}
&&A(r)=r^2 + c_1 + r c_2 - 2 \alpha (r-r') \theta(r-r').\label{A1}
\end{eqnarray}
here $\theta(x)$ is the Heaviside theta (step function), 
\begin{eqnarray}
\label{green33}
\theta(r-r') =\left\{	\begin{array}{ll}  0  & \mbox{if } r<r' \\			1 & \mbox{if } r>r'
\end{array}\right.
\end{eqnarray}
One can directly show that the above metric function solves ODE given in eq. (\ref{ODEA}), because as we know  $(x-x')\delta(x-x')=0$, as a result $2 \delta(x-x')+(x-x')\frac{d\delta(x-x')}{dx}\equiv \delta(x-x')$.  The metric explicitly written as 
\begin{eqnarray}
ds^2=\left\{	\begin{array}{ll}  (r^2 + c_1 + r c_2)dt^2+\frac{dr^2}{r^2 + c_1 + r c_2}  & \mbox{if } r<r' \\		(r^2 + c_1 + r c_2-2\alpha(r-r'))dt^2+\frac{dr^2}{r^2 + c_1 + r c_2-2\alpha(r-r')} & \mbox{if } r>r'
\end{array}\right.
\end{eqnarray}
The metric suffered from a first jumping singularity at $r=r'$, where the metric radial derivative $\partial_r g_{\mu\nu}|_{r=r'-0}\neq \partial_r g_{\mu\nu}|_{r=r'+0}$. It can be interpreted as a discontinuity in the connections. For the exterior region $r>r'$, we observe that if we apply the following transformation to the metric within the interior region $r<r'$, the metrics $ds^2_{r>r'}$ will be mapped to the one on the exterior region,
\begin{eqnarray}
&&c_1\to c_1+2\alpha r',\ \ c_2\to c_2-2\alpha 
\end{eqnarray}
consequently we deduce that the exterior  metric can be mapped analytically to the interior metric as the following 
\begin{eqnarray}
&&ds^2(r<r')_{\Big[c_1\to c_1+2\alpha r',\ \ c_2\to c_2-2\alpha \Big]}\longrightarrow ds^2(r>r')
\end{eqnarray}
The singularity sphere with equation $r=r'$ can be identified only with a unique metric for inner or outer regions. 
\par
The metric function has the following regular form for both regions but with different values of the metric parameters $(c_1,c_2)$
\begin{eqnarray}
&&g_{tt}=g_{rr}^{-1}=r^2 + c_1 + r c_2
\end{eqnarray}
The metric can be considered as a blackhole with  horizon radius $r_h$
\begin{eqnarray}
&&r_h=\frac{1}{2}(-c_2+\sqrt{c_2^2-4c_1}), \ \ c_1<0,c_2>0.
\end{eqnarray}
The second root of the equation $g_{tt}(r)=0$ is negative and completely will be removed from further considerations. Note that one can study near horizon geometry , we have 
\begin{eqnarray}
&&A(r)=\sqrt{c_2^2-4c_1}(r-r_h)+\mathcal{O}((r-r_h)^2)
\end{eqnarray}
We define a new coordinate $z=r-r_h$, using this coordinate the near horizon geometry represented as the following,
\begin{eqnarray}
&&ds^2\approx \frac{4}{\sqrt{c_2^2-4c_1}}\Big(dz^2+(\frac{c_2^2-4c_1}{4}) z^2dt^2
\Big)
\end{eqnarray}
The metric remains smooth in the vicinity of the $z=0$, if and only if 
\begin{eqnarray}
&&t\cong t+\frac{4\pi}{\sqrt{c_2^2-4c_1}}
\end{eqnarray}
It defined the temperature as $T=\frac{\sqrt{c_2^2-4c_1}}{4\pi}$.

%%%%%%%%%%%%%%%%%%%%%%%%%
\par
For the case with the deficit parameters more than one, when  $\alpha_{I}\equiv (\alpha_1,\alpha_2,..,\alpha_k)$ and using the metric form as we used in the case with a single deficit parameter $\alpha$, 
the Witten's equation  for the deformed hyperboic geometry changes to the following form,
\begin{eqnarray}
&&A''=2(1-\sum_{i=1}^{k}\alpha_i\delta^{(2)}(x-x'_i)).
\end{eqnarray}
We are lucky to have a linear ODE, using the superposition principle 
 , exact solution for a set of the deficit parameters and singularities can be obtained using   the general solution which we obtained for the singular parameter i.e, eq. (\ref{A1}), 
 as $A_{k}(r)=\sum_{i=1}^k A_i$ where
\begin{eqnarray}
&&A_i= \frac{r^2}{k}+ c_{i}^{(1)} + rc_{i}^{(2)} - 2\alpha_i(r-r'_i) \theta(r-r'_i)
\end{eqnarray}
The general metric solution for $\mathcal{M}_{g,k}$ is
\begin{eqnarray}
&&A_{k}(r)= r^2+\sum_{i=1}^{k}\Big( c_{i}^{(1)} + rc_{i}^{(2)} + 2\alpha_i(r'_i-r) \theta(r-r'_i)\Big).
\end{eqnarray}
It can be understood as a superposition of the single mode solutions as one of the rare cases where the Einstein gravity still respects superposition principle.  The metric explicitly written as 
\begin{eqnarray}
ds_{k}^2=\left\{	\begin{array}{ll}  ( r^2+\sum_{i=1}^{k}\Big( c_{i}^{(1)} + rc_{i}^{(2)} \Big) )dt^2+\frac{dr^2}{ r^2+\sum_{i=1}^{k}\Big( c_{i}^{(1)} + rc_{i}^{(2)}\Big)}  & \mbox{if } r<r' \\	
(  r^2+\sum_{i=1}^{k}\Big( c_{i}^{(1)} + rc_{i}^{(2)} + 2\alpha_i(r'_i-r) \Big) )dt^2+\frac{dr^2}{  r^2+\sum_{i=1}^{k}\Big( c_{i}^{(1)} + rc_{i}^{(2)} + 2\alpha_i(r'_i-r) \Big)
} & \mbox{if } r>r'
\end{array}\right.
\end{eqnarray}
In the above metric, the set of the integration parameters 
\begin{eqnarray}
&&c_{i}^{(1)}\to c_{i}^{(1)}+2\alpha r',\ \ c_{i}^{(2)}\to c_{i}^{(2)}-2\alpha 
\end{eqnarray}
under the following transformations the metrics for interior and exterior regions remains unaltered,
\begin{eqnarray}
&&ds_{k}^2(r<r')_{\Big[c_{i}^{(1)}\to c_{i}^{(1)}+2\alpha_i r',\ \ c_{i}^{(2)}\to c_{i}^{(2)}-2\alpha_i  \Big]}\longrightarrow ds_{k}^2(r>r')
\end{eqnarray}
The near horizon geometry and temperature can be recovered as we obtained in the case with a single deficit parameter only if we replace $c_1,c_2$ as the following,
\begin{eqnarray}
&&c_1\to \sum_{i=1}^{k}c_{i}^{(1)},
\ \ c_2\to \sum_{i=1}^{k}c_{i}^{(2)}.
\end{eqnarray}
The temperature for the case with many finite deficit parameters can be written as the following expression,
\begin{eqnarray}
&&T=\frac{\sum_{i,j=1}^{k}c_{i}^{(2)}c_{j}^{(2)}-4\sum_{i=1}^{k}c_{i}^{(1)}}{4\pi}
\end{eqnarray}
where we assumed that $\sum_{i=1}^{k}c_{i}^{(1)}<0$. 
%%%%%%%%%%%%%%%%%
\section{On singular manifold with  time-dependent metrics }
In the previous section we investigated exact blackhole  solutions when the metric considered as time independent, i.e, when $A=A(r)$. Under this assumption we integrate a linear ODE with Dirac delta source term and the analysis could be extended to the situation when  the deficit parameters $\alpha_i$ be more than one but still remains finite. In this section we want to see if we relax the constraint of time independence and allow the metric to be time dependent, how the singular surfaces will be deformed adequately. Firstly we note that although in two dimensions it is possible to reduce the metric to a conformal flat Euclidean  metric in general, but when the metric is considered time dependent,  
\begin{eqnarray}
&&ds^2=g_{tt}(t,r)dt^2+2g_{tr}(t,r)dtdr+g_{rr}(t,r)dr^2
\end{eqnarray}
the metric can't not be written as a conformal flat form. Let us see what is going to happen if one factorize the above arbitrary time dependent metric,
\begin{eqnarray}
&&ds^2=g_{tt}(t,r)\Big(dt+A_{r}(t,r)dr
\Big)^2+\Big(g_{rr}(t,r)-\frac{g_{tr}(t,r)^2}{g_{tt}(t,r)}\Big)dr^2
\end{eqnarray}
where the non static vector potential (or angular velocity in the terminology of the lower dimensional blackholes) is defined as $A_{r}(t,r)\equiv \frac{g_{tr}(t,r)}{g_{tt}(t,r)}$. If one needs a conformal flat form of the above metric, we are rquested to find a set of the new coordinates $T,R$ such that ,
\begin{eqnarray}
&&dT(t,r)=dt+A_{r}(t,r)dr\label{eqT}
\\&&
dR(t,r)=\sqrt{g_{rr}(t,r)-\frac{g_{tr}(t,r)^2}{g_{tt}(t,r)}}dr\label{eqR1}
\end{eqnarray}
The first equation is not a regular Pfaffian form. Let us try to make it Pfaffian using an auxiliary function $\mu(t,r)$, where 
\begin{eqnarray}
&&\frac{\partial \mu(t,r)}{\partial r}=\frac{\partial \Big(\mu(t,r)A_{r}(t,r)\Big)}{\partial t}
\end{eqnarray}
The aim is to find at least one  auxiliary function $\mu(t,r)$ to make  (\ref{eqT}) a Pfaffian  form and makes us possible  to integrate (\ref{eqR1}). After a carefully investigation, we see that the following cases are possible:
\begin{itemize}
    \item $\mu(t,r)=\mu(t)$: Following this constraint we obtain $\mu(t)A_{r}(t,r)=C(r)$ and finally it suggests that $\mu(t)=\frac{C(r)}{A_{r}(t,r)}$. The only possibility is when $A_{r}(t,r)=A_1(r)A_{2}(t)$. With this choice of the vector potential, we can show that 
    \begin{eqnarray}
    &&\frac{g_{tr}(t,r)}{g_{tt}(t,r)}=\frac{C(r)}{A_1(r)A_{2}(t)}=\frac{\tilde{C}(r)}{A_2(t)}
    \end{eqnarray}
    under these constraints, the first  (\ref{eqT}) converts to  a Pfaffian  form ,
    \begin{eqnarray}
&&dT(t,r)=\mu(t)dt+C(r)dr=d\Phi(t,r)
\end{eqnarray}
    where the potential function 
        \begin{eqnarray}
&&\Phi(t,r)=\tilde{C}(r)\int \frac{dt}  {A_{2}(t)} +\int dr C(r)
   \end{eqnarray} 
    then the second integral gives us
  \begin{eqnarray}
&&
dR(t,r)=dr\sqrt{g_{rr}(t,r)-(\frac{\tilde{C}(r)}{A_2(t)})^2g_{tt}(t,r)}\label{eqR}
\end{eqnarray}  
    the above differential form can be integrated only if $g_{rr}(t,r)-(\frac{\tilde{C}(r)}{A_2(t)})^2g_{tt}(t,r)=B(r)$. Using the above set of the new coordinates $T,R$, the metric reduces to a conformal flat form after defining a set of the null coordinates $U,V$ adequately(see for example \cite{Momeni:2020zkx}).
    
    \item $\mu(t,r)=\mu(r)$: In this situation, the factor $\mu(r)$ can be obtained as
      \begin{eqnarray}
&&\mu(r)=\exp{\frac{\partial }{\partial t}}\int dr A_{r}(t,r)
\end{eqnarray} 
it becomes meaningful if and only if $A_{r}(t,r)\propto t $. Within this condition we obtain $\mu(r)=\exp{\int dr A_{r}(r)}$. This case is also provides set of the appropriate null coordinates. But as we checked , in general one can't reduce any two dimensional metric  to a flat conformal form. 
\end{itemize}
In dJT, the singular surfaces with time dependent metric are completely different from the static ones. Let us simple consider a very restricted form of such metrics when  the metric is given by a single gauge $A(t,r)=g_{tt}=g_{rr}^{-1},\ \ g_{tr}=0.$ Even if we consider the case with one deficit parameter $\alpha$, the Witten's equation for the singular metrics reduces to a nonlinear second order PDE, with a source term as $\delta(r-r')\delta(t-t')$. The model can be considered as a nonlinear wave equation in an inhomogeneous medium. There won't be an easy exact solution for the metric function $A(t,r)$, that means one can't construct (easily) an analytic Green function for such  wave equation in such an inhomogeneous medium. A further study can be done if we provide an appropriate set of the initial conditions for the Green function and integrate numerically the Green function very carefully in the vicinity of the singularity point. Otherwise , we have to find Green function perturbative , order by order if any dimensionless perturbation parameter can be found in the model

%%%%%%%%%%%%%%%%%%%%%%%

%%%%%%%%%%%%%%%%
\section{Conclusion}
In this work, we investigated deformed geometries for deformed JT gravity recently proposed by E.\,Witten \cite{Witten:2020ert}. As claimed by Witten, there should not be any type of classical geometry for deficit angle $\alpha>2\pi$ . We found a class of metric solutions in the null coordinate as a non trivial solution for the nonlinear PDE formulated as Witten's problem. The metric  expressed as the Green function for the operator. There is a discontinuity in the metric derivative and that implies a type of the phase transition between AdS and dS metrics. Although Witten's equation for the deformed hypersurface with $\alpha\neq 0$ is nonlinear, we demonstrated that some linear independence still remains in theory. As an attempt to prove the difference between two independent exact solutions to the homogeneous equation, we studied the null geodesics of two metrics. It has been shown that the trajectory of the test particle coincides with the classical trajectories in two different patches of the AdS. Although we didn't compute the free energy, a qualitative discussion provides more evidence about a coexistence phase of both patches of the AdS in dJT. It will  be very interesting to compare such formal phase transition with the realistic description which has been investigated in the details is Re. \cite{Witten:2020ert}. The thermodynamical phase transition proposed for dJT in the above reference is based on the study of the free energy of a given exact dilatonic blackhole obtained in a suitable scale invariant (gauge fixed) Euclidean form for the metric. In our study a pair of the patches for the AdS spacetime are surprisingly appeared in the model as  exact solutions (with a signature change) to the homogeneous Green equation. The phase transition from static patch of the Ads to non static one which we proposed in our work can be interpreted as a simple phase transition from an Euclidean AdS to another Euclidean AdS along a signature change.e,
\begin{eqnarray}
&&AdS\Longrightarrow_{Signature-changed } AdS
\end{eqnarray}
Because the signature change in the Euclidean (or even Lorentzian) metrics usually occurs when we pass a horizon (or getting close to the singularity), the phenomena here looks a bit odd and we have to study it in more details in out forthcoming work(s). One possible description could be as the following: small quantum fluctuations in the dJT could makes a clouds surrounding the AdS blackhole. Although it can't make a horizon , it is possible to make a transitive horizon with very short life time. Consequently the metric in the outer region of the cloud described by the usual Euclidean AdS while one passes the temporal cloudy horizon, the metric signature eventually changes. 
We mention here that the phase transition described in a purely classical sense using the classical trajectories. We expanded our study by considering blackhole solutions in theory with a single and finite number of the deficit parameters $\alpha$. In the case of the single parameter we found a class of exact static, spherically symmetric two dimensional blackhole solutions with a non zero temperature and a well behaviour horizon. We show that the metric for the  exterior region $r>r'$ beyond the singularity point $r=r'$ can be mapped smoothly to the interior metric within the region $r<r'$ if one apply a class of simple algebraic transformations between the set of the integration constants $c_1,c_2$. When we have more than one deficit parameter, we showed that the equation for the singular surfaces remains linear . Using the superposition principle we construct the finite $k$ metric from $k$ copies of the single singular  metric. By constraining the parameters of the metric we show that the horizon remained thermal with a similar temperature for the single hole metric. 

%%%%%%%%%%%%%%%%%

\end{document}